\documentclass[12pt,aps]{revtex4}
\usepackage{amsmath}
\usepackage{amssymb}
\usepackage{graphicx}
\usepackage{epsfig}

\newcommand{\beq}{\begin{equation}}
\newcommand{\eeq}{\end{equation}}
\newcommand{\bea}{\begin{eqnarray}}
\newcommand{\eea}{\end{eqnarray}}

\begin{document}

\title{Coupled-channel effective field theory and proton-$^7$Li 
scattering}
\author{Vadim Lensky}\thanks{On leave from Institute for Experimental and Theoretical Physics, B.~Cheremushkinskaya~25,
117218, Moscow, Russia}
\author{Michael C.~Birse}
\affiliation{Theoretical Physics Group, School of Physics and Astronomy,
The University of Manchester, Manchester, M13 9PL, UK\\}

\begin{abstract}
We apply the renormalisation group (RG) to analyse scattering by short-range forces in systems with coupled channels.
For two $S$-wave channels, we find three fixed points, corresponding to systems with zero, one or two bound or virtual
states at threshold. We use the RG to determine the power countings for the resulting effective field theories.
In the case of a single low-energy state, the resulting theory takes the form of an effective-range expansion
in the strongly interacting channel. We also extend the analysis to include the effects of the Coulomb interaction between charged 
particles. The approach is then applied to the coupled $p+{^7}$Li and $n+{^7}$Be channels which couple to a $J^P=2^-$ state of $^8$Be
very close to the $n+{^7}\mbox{Be}$ threshold. At next-to-leading order, we are able to get a good description of the $p+{^7}$Li phase shift
and the $\mathrm{{^7}Be}(n,p)\mathrm{{^7}Li}$ cross section using four parameters. Fits at one order higher are similarly good but 
the available data are not sufficient to determine all five parameters uniquely. 

\end{abstract}
\maketitle

\section{Introduction}\label{sec:Intro}

Effective field theories (EFTs) now provide a widely used tool in studies of
a variety of systems in nuclear, particle, and atomic physics
(for reviews, see Refs.~\cite{bvkrev,bhrev,eprev}). The idea behind 
them is that, in order to describe processes at sufficiently small energies, one 
does not need to treat explicitly the underlying short-distance physics. 
Provided there is good separation between the energy scales of interest and 
those of the short-distance physics, one can work with the appropriate low-energy
degrees of freedom and expand the resulting theory in powers of ratios
of low-energy to high-energy scales. A key ingredient of any such theory is thus 
the ``power counting" used to organise this expansion of its interactions.

Some of the main applications of these ideas have been to few-nucleon systems on 
momentum scales of the order of the pion mass. These make use of Chiral Perturbation
Theory to determine the long-range pion-exchange forces between nucleons, and 
replace the unresolved short-range forces by contact interactions. At lower
energies, even pion-exchange forces are not resolved and one can work instead 
with a simpler, pionless EFT, involving only contact interactions~\cite{vk99}.
In nucleon-nucleon $S$-waves, the deuteron bound state and the $^1S_0$
virtual state have the effect of enhancing the low-energy wave functions at short
distances. The resulting power-counting in these channels shows a strong
promotion of the leading contact interactions, which must be treated 
nonperturbatively. 

More recently, a similar approach, also based on contact interactions, has been applied 
to larger nuclei which have low-energy states or resonances with a clustered 
structure. This theory, known as ``halo EFT", has been applied to various weakly 
bound nuclear systems~\cite{bhvk02,bhvk03,hhvk08,rh11,hp11}. A number of these
applications are to systems of astrophysical importance, and some of the most
interesting of these involve reactions whose cross sections are enhanced by 
resonant or virtual states lying very close to a threshold. EFTs that incorporate 
such states have been developed in Refs.~\cite{cgvk04,g09}.

Theories describing coupled channels, such as those developed by Cohen 
\textit{et al.}~\cite{cgvk04} are needed for applications to reactions. 
Here we use the renormalisation group (RG) to determine the possible power countings
for systems with two scattering channels. This is an extension of the 
Wilsonian approach~\cite{Wilson:1973jj}, introduced in Ref.~\cite{Birse:1998dk} to study single-channel 
systems with short-range interactions. We first determine the fixed points of the 
RG, and then use the linearised RG equation to analyse the scale dependences of 
perturbations around these points and hence find the power counting.

In the case of two coupled channels, we find three fixed points.
One of these is just the trivial one, describing a non-interacting system.
This is the appropriate starting point for an EFT describing a weakly interacting system,
with no enhancements of the scattering near thresholds. A second has strong interactions
in both channels, corresponding to two virtual states lying at the higher threshold.
This corresponds to the case considered in Ref.~\cite{cgvk04}, where the matrices of energy-independent 
interactions are promoted compared to simple dimensional power counting. 

Lastly there is a fixed point with a single virtual state at the threshold.
This is the one of most physical relevance since it requires ``fine-tuning" of only
one quantity. This is in contrast to EFT studied in Ref.~\cite{cgvk04}
where two parameters, the eigenvalues of the leading matrix interaction, 
must be fine-tuned. Both that EFT and the one studied here correspond to 
coupled-channel extensions of the effective-range expansion~\cite{Bethe:1949yr,rs61,bkps82}. 
They provide potential alternatives to the $R$-matrix approach~\cite{lt58} 
that is widely used to analyse reactions with resonant or virtual states~\cite{daacvf04}. 

As well as being applicable to low-energy resonances in nuclear physics, the same 
coupled-channel EFT can describe near-threshold states in other contexts, such as
quarkonium states in hadronic physics. One particularly interesting example is the 
$X(3872)$~\cite{belle03}, which lies very close to the $D^0\bar D^{*0}$ threshold. 
This was recently studied by Hanhart \textit{et al.}~using a phenomenological
coupled-channel approach~\cite{hkn11}. It would be very interesting to examine it 
within the EFT framework developed here.

In general, at least one of the channels in a nuclear reaction involves two charged 
particles. As well as the short-ranged nuclear forces, we therefore need to treat 
the Coulomb interaction. EFT and RG techniques have been developed to analyse scattering 
in the presence of a Coulomb potential~\cite{kr99,Barford:2002je,Ando:2007fh,ab08,ab08_1}. These show
that the $1/r$ singularity is not strong enough to alter basic power counting, which 
is still that of an effective-range expansion.
However there is a new low-energy scale $\kappa$, the inverse Bohr radius.
The interaction also leads to a logarithmic divergence that needs to be 
renormalised by a counterterm linear in $\kappa$. This logarithmic behaviour makes 
it impossible to disentangle the purely strong-interaction scattering length from 
the scatttering data. In the present work, we extend these treatments to coupled-channel
systems, concentrating on the case where only one of the channels consists of two charged 
particles.

As a practical illustration of the resulting EFT, we apply it to the system of coupled 
$p+{^7}$Li and $n+{^7}$Be channels. The reaction $^7$Be$(n,p)$$^7$Li is an important one 
in the context of primordial nucleosynthesis of $^7$Li. However, it is one that has been 
well studied experimentally over many years~\cite{PhysRev.109.105,NuclPhysA.206.353,PhysRevC.37.917,JPhysG.29.395,cvfdaa04}
and so it is not expected to provide an explanation for the observed abundance of $^7$Li~\cite{cfo08}.
The astrophysical importance of this reaction is a consequence of its enhancement by
a virtual $2^-$ state of $^8$Be lying very close to the $n+{^7}$Be threshold. This makes
the system an ideal one to test our EFT approach, as it has a single low-energy state
which is coupled strongly to both physical channels. We calculate scattering at next-to-next-to-leading
order (NNLO) in our power couning and determine the corresponding low-energy parameters 
from the available experimental data. We also examine the nature of ${^8}$Be pole
within our approach.

The structure of our paper is as follows. In Sec.~\ref{sec:RG} we use the RG to 
analyse scattering in systems with two coupled channels. The extension of this to 
systems with Coulomb interactions is outlined in the Appendix. In Sec.~\ref{sec:CC} we 
apply the resulting EFT to the example of coupled $p+{^7}$Li and $n+{^7}$Be channels.
We close with some conclusions in Sec.~\ref{sec:Conclusion}.

\section{RG Analysis}\label{sec:RG}

\subsection{Short-range forces}

Here we use the RG to analyse scattering in a two-channel system with only
short-range forces. The Lippmann-Schwinger equation for the $T$-matrix can be written
schematically as 
\begin{equation}
\mathbf{T}=\mathbf{V}+\mathbf{V}\mathbf{G}\mathbf{T},
\end{equation}
where 
\begin{equation}
\mathbf{G}=\left(\begin{array}{cc}
G_1 & 0 \cr
0 & G_2 \end{array}\right)
\end{equation}
in terms of the free Green's functions for the individual channels.
These can be written in momentum space as
\begin{equation}
G_i=2M_i\int \frac{\mathrm{d}^3\vec q}{2\pi^3}\,
\frac{1}{p_i^2-q^2+\mathrm{i}\,\epsilon},
\end{equation}
where $M_i$ is the reduced mass in channel $i$ and $p_i$ is the on-shell
momentum. Taking zero energy to be the lower threshold and $\Delta$ to be
the threshold energy of the second channel, we have
\begin{eqnarray}
p_1&=&\sqrt{2M_1E}\equiv p,\\
\noalign{\vspace{5pt}}
p_2&=&\sqrt{2M_2(E-\Delta)}=\sqrt{\frac{M_2}{M_1}\left(p^2-\delta^2\right)},
\end{eqnarray}
where the momentum scale $\delta$ associated with the difference between the thresholds
is
\begin{equation}
\delta=\sqrt{2M_1\Delta}.
\end{equation}
Specialising to the case of two $S$-wave channels, we can take the short-range
potential to consist simply of $\delta$-functions with energy-dependent 
coefficients. This is because terms that depend on the off-shell momenta 
(that is, derivatives of $\delta$-functions) are of the same or higher order as 
the corresponding energy-dependent ones and so are not essential for a description
of the on-shell scattering \cite{Birse:1998dk,ab08,ab08_1}. This leaves the system with 
two low-energy scales: $p$ and $\delta$. These provide the expansion parameters 
of our EFT.

For $S$-wave channels with contact interactions, the coupled integral
equations for $\mathbf{T}$ reduce to algebraic equations for the coefficients of 
the $\delta$-functions, 
\begin{equation}
\mathbf{T}(p,\delta)=\mathbf{V}(p,\delta)
+\mathbf{V}(p,\delta)\mathbf{J}(p,\delta)
\mathbf{T}_(p,\delta),
\end{equation}
where the diagonal matrix $\mathbf{J}(p,\delta)$ consists of
the loop integrals
\begin{equation}
J_i(p,\delta)=2M_i\int \frac{\mathrm{d}^3\vec q}{2\pi^3}\,
\frac{1}{p_i^2-q^2+\mathrm{i}\,\epsilon}.
\label{eqn:LSeq}
\end{equation}
These loop integrals are divergent and so need to be regularised in some way. In
Ref.~\cite{Birse:1998dk}, a sharp momentum cutoff was used. The resulting expressions
for the potentials are a little cumbersome and so here we use dimensional 
regularisation with the power divergence subtraction scheme intoduced by Kaplan, 
Savage and Wise \cite{Kaplan:1998,Kaplan:1998_1}. As discussed in Ref.~\cite{Birse:1998dk}, the 
results for ``universal" quantities --- RG fixed points and power countings ---
are the same as those obtained with a cutoff. 

In the absence of Coulomb interactions, there are only linear divergences and 
subtracting these at the scale $\mu$ gives
\begin{equation}
J_i(p,\delta,\mu)=-\,\frac{M_i}{2\pi}\,(\mu+\mathrm{i}\,p_i).
\end{equation}
The subtraction scale $\mu$ is arbitary and so the physical scattering amplitudes
in $\mathbf{T}$ should be independent of it. To cancel the $\mu$ dependence of the 
loop integrals in Eq.~(\ref{eqn:LSeq}), the potential matrix $\mathbf{V}(p,\delta,\mu)$ 
must run with $\mu$ according to 
\begin{equation}
\frac{\partial\mathbf{V}}{\partial\mu}=-\mathbf{V}\,
\frac{\partial\mathbf{J}}{\partial\mu}\,\mathbf{V}.
\label{eqn:Veq}
\end{equation}

To determine the possible power countings, we need to put Eq.~(\ref{eqn:Veq}) into the form 
of a standard RG equation. We do this by defining dimensionless variables corresponding 
to the low-energy scales,
\begin{equation}
\hat p=p/\mu,\quad \hat\delta=\delta/\mu,
\end{equation}
and a rescaled potential,
\begin{equation}
\hat{\mathbf{V}}=\frac{\mu}{2\pi}\,\mathbf{M}^{1/2}\,\mathbf{V}\,\mathbf{M}^{1/2},
\end{equation}
where
\begin{equation}
\mathbf{M}=\left(\begin{array}{cc}
M_1 & 0 \cr
0 & M_2 \end{array}\right).
\end{equation}
The evolution equation then becomes
\begin{equation}
\mu\,\frac{\partial\hat{\mathbf{V}}}{\partial\mu}
=\hat p\,\frac{\partial\hat{\mathbf{V}}}{\partial\hat p}
+\hat\delta\,\frac{\partial\hat{\mathbf{V}}}{\partial\hat\delta}
+\hat{\mathbf{V}}+\hat{\mathbf{V}}^2.
\label{eqn:RGeq}
\end{equation}

The dimensionless form of this equation allows us to look for fixed-point solutions 
which describe scale-free physical systems. We then expand the potential around one 
of these points and find perturbations that scale with definite powers of the regulator 
scale $\mu$. Because of the rescaling, this power of $\mu$ counts the net power of 
low-energy scales in the corresponding term in the potential and so gives the order of
that term in the power counting \cite{Birse:1998dk}. More precisely, a term that scales 
as $\mu^\nu$ is of order $Q^{\nu-1}$, where $Q$ denotes a generic low-energy scale.

One obvious fixed point is the trivial solution to Eq.~(\ref{eqn:RGeq}): 
$\hat{\mathbf{V}}=0$. The EFT based on the expansion around this point decribes systems
that interact weakly at low energies. The leading (energy-independent) term in the potential
scales as $\mu^1$ and so is of order $Q^0$. Each power of the energy ($p^2$) and splitting ($\delta^2$)
increases the order by $Q^2$ as expected from naive dimensional analysis (or ``Weinberg power counting")~\cite{w90,w90_1}.

A second, nontrivial fixed point can easily be constructed by analogy with that in the 
single-channel case. If we express the RG equation in terms of $\hat{\mathbf{V}}^{-1}$,
then it takes the simpler, linear form,
\begin{equation}
\mu\,\frac{\partial}{\partial\mu}\hat{\mathbf{V}}^{-1}
=\hat p\,\frac{\partial}{\partial\hat p}\hat{\mathbf{V}}^{-1}
+\hat\delta\,\frac{\partial}{\partial\hat\delta}\hat{\mathbf{V}}^{-1}
-\hat{\mathbf{V}}^{-1}-\mathbf{I},
\label{eqn:RGeqVinv}
\end{equation}
where $\mathbf{I}$ is the $2\times 2$ identity matrix. 
This has the $\mu$-independent solution
\begin{equation}
\hat{\mathbf{V}}_0=-\mathbf{I},
\label{eqn:Vtwobound}
\end{equation}
which is just the $2\times 2$-matrix equivalent of the fixed-point that corresponds to
the effective-range expansion for a single channel \cite{Birse:1998dk}. It describes 
a system with bound states at threshold in both channels.

Since Eq.~(\ref{eqn:RGeqVinv}) is linear, exact solutions can be constructed and 
expressed in terms of perturbations that scale with definite powers of $\mu$:
\begin{equation}
\hat{\mathbf{V}}(\hat p,\hat\delta,\mu)^{-1}=-\mathbf{I}
-\sum_{n,m}\mathbf{C}_{nm}
\,\mu^{2n+2m-1}\,\hat p^{2n}\,\hat \delta^{2m}.
\label{eqn:Vinvgen}
\end{equation}
This shows that the fixed point has two unstable directions, given by
the two eigenvectors of $\mathbf{C}_{00}$. The expansion is the appropriate one 
for systems where both eigenvalues of $\mathbf{C}_{00}$, denoted by $c_{1,2}$, are 
unnaturally small, corresponding to two low-energy bound or virtual states. This is 
the counting considered by Cohen \textit{et al.}~\cite{cgvk04}, who treated all elements 
of the energy-independent matrix $\mathbf{C}_{00}$ as leading-order in their expansion.
In this case we can keep $\mu>c_{1,2}$ so that the constant $-1$ provides the leading 
term in $\hat{\mathbf{V}}^{-1}$. Taking the inverse, we get the potential in the form
\begin{equation}
\hat{\mathbf{V}}(\hat p,\hat\delta,\mu)=-\mathbf{I}
+\sum_{n,m}\mathbf{C}_{nm}
\,\mu^{2n+2m-1}\,\hat p^{2n}\,\hat \delta^{2m}+\cdots,
\label{eqn:Vstrong}
\end{equation}
which shows that all terms are enhanced by two orders compared to Weinberg power 
counting.

Of more interest is a second nontrivial fixed point, which has the single-term
separable structure
\begin{equation}
\hat{\mathbf{V}}=\mathbf{u}\,\hat V_0(\hat p, \hat\delta)\,\mathbf{u}^{\dagger},
\label{eqn:Vseparable}
\end{equation}
where
\begin{equation}
\mathbf{u}=\left(\begin{array}{c}
\cos\phi \cr
\sin\phi\end{array}\right).
\end{equation}
Its strength $\hat V_0(\hat p, \hat\delta)$ satisfies
\begin{equation}
\hat p\,\frac{\partial\hat V_0}{\partial\hat p}
+\hat\delta\,\frac{\partial\hat V_0}{\partial\hat\delta}
+\hat V_0+\hat V_0^2=0,
\end{equation}
and hence is just 
\begin{equation}
\hat V_0(\hat p, \hat\delta)=-1.
\end{equation}
This potential generates a single zero-energy bound state whose overlaps
with the two scattering channels are given by the components of $\mathbf{u}$. 
In the orthogonal combination of channels, there is no interaction.

Physical systems with a single low-energy bound or virtual state can be described 
by potentials close to this fixed point. Expanding $\hat{\mathbf{V}}$ around it 
and linearising the RG equation, one can show that interactions in the channel 
$\mathbf{u}$ are enhanced by two orders, as in Eq.~(\ref{eqn:Vstrong}) above.
In contrast, those in the orthogonal channel, specified by
\begin{equation}
\mathbf{v}=\left(\begin{array}{c}
-\sin\phi \cr
\cos\phi\end{array}\right),
\end{equation}
have natural coefficients and can be organised according to Weinberg's
power counting. Terms in the potential that couple the $\mathbf{u}$ and
$\mathbf{v}$ channels are enhanced by one order.

In practice, an easier way to construct the expansion of a potential around 
this point is to start from the general solution to the RG equation,
Eq.~(\ref{eqn:Vinvgen}). If only one of the eigenvalues, $c_1$, of 
$\mathbf{C}_{00}$ is unnaturally small, we should choose our renormalisation 
scale $\mu$ such that $c_1<\mu\ll c_2$. In this case, the term $c_2/\mu$ 
becomes the dominant one in the expansion of $\hat{\mathbf{V}}^{-1}$. As a result, 
the expansion of $\hat{\mathbf{V}}$ has a different structure from the case just 
discussed, Eq.~(\ref{eqn:Vstrong}). In particular, the expansions in the channels corresponding to the two 
eigenvectors of $\mathbf{C}_{00}$ have different power countings \footnote{If both 
eigenvalues of $\mathbf{C}_{00}$ are of natural size ($c_{1,2}\gg\mu$), then the whole 
term $\mathbf{C}_{00}/\mu$ is the largest one in $\hat{\mathbf{V}}^{-1}$. Inverting to 
get $\hat{\mathbf{V}}$ itself, we see that the leading term, $\mathbf{C}_{00}^{-1}\mu$
is of order $\mu$, as it should be since this is an expansion around the trivial 
fixed point $\hat{\mathbf{V}}=0$.}.

To find the corresponding expansion of the whole potential, we need to invert 
Eq.~(\ref{eqn:Vinvgen}). It is convenient to introduce four matrices 
$\mathbf{P}_\alpha$, defined by
\begin{eqnarray}
\mathbf{P}_\mathrm{u}=\mathbf{u}\,\mathbf{u}^\dagger,\quad
\mathbf{P}_\mathrm{v}=\mathbf{v}\,\mathbf{v}^\dagger,\quad
\mathbf{P}_\mathrm{m}=\mathbf{u}\,\mathbf{v}^\dagger,\quad
\mathbf{P}_\mathrm{m}^\dagger=\mathbf{v}\,\mathbf{u}^\dagger,
\end{eqnarray}
where $\mathbf{u}$ and $\mathbf{v}$ denote the eigenvectors of $\mathbf{C}_{00}$ 
corresponding to the eigenvalues $c_1$ and $c_2$ respectively. In terms of these,
we can write the general solution for the inverse potential, Eq.~\eqref{eqn:Vinvgen},
in the form
\beq
\hat{\mathbf{V}}(\hat p,\hat\delta,\mu)^{-1}=
\hat{f}_\mathrm{u}(\hat p,\hat\delta,\mu)^{-1}\mathbf{P}_\mathrm{u}+
\hat{f}_\mathrm{v}(\hat p,\hat\delta,\mu)^{-1}\mathbf{P}_\mathrm{v}+
\hat{f}_\mathrm{m}(\hat p,\hat\delta,\mu)^{-1}\mathbf{P}_\mathrm{m}+
\hat{f}_\mathrm{m}^{*}(\hat p,\hat\delta,\mu)^{-1}\mathbf{P}_\mathrm{m}^\dagger,
\label{eqn:Vi1app}
\eeq
where the functions $\hat{f}_\alpha(\hat p,\hat\delta,\mu)^{-1}$ are defined by
\bea
\hat{f}_\mathrm{u}(\hat p,\hat\delta,\mu)^{-1}&=&-1-\sum_{n,m}c_{nm}^{(\mathrm{u})}\,
\mu^{2n+2m-1}\,\hat p^{2n}\,\hat \delta^{2m},\\
\hat{f}_\mathrm{v}(\hat p,\hat\delta,\mu)^{-1}&=&-1-\sum_{n,m}c_{nm}^{(\mathrm{v})}\,
\mu^{2n+2m-1}\,\hat p^{2n}\,\hat \delta^{2m},\\
\hat{f}_\mathrm{m}(\hat p,\hat\delta,\mu)^{-1}&=&-\sum_{n,m}c_{nm}^{(\mathrm{m})}\,
\mu^{2n+2m-1}\,\hat p^{2n}\,\hat \delta^{2m},
\eea
and $c_{00}^{(\mathrm{u})}=c_1,\ c_{00}^{(\mathrm{v})}=c_2$. Note that $c_{00}^{(\mathrm{m})}=0$.
Provided no other channels are open, this potential should be Hermitian, with $c_{nm}^{(\mathrm{u,v})}$ real. 

Inverting Eq.~\eqref{eqn:Vi1app} we get the potential,
\beq
\hat{\mathbf{V}}=\left[f_\mathrm{v}^{-1}\mathbf{P}_\mathrm{u}+f_\mathrm{u}^{-1}\mathbf{P}_\mathrm{v}-f_\mathrm{m}^{-1}\mathbf{P}_\mathrm{m}-f_\mathrm{m}^{*-1}\mathbf{P}_\mathrm{m}^\dagger\right]
[f_\mathrm{u}^{-1}f_\mathrm{v}^{-1}-f_\mathrm{m}^{-1}f_\mathrm{m}^{*-1}]^{-1}.
\eeq
For the situation of interest, we need to expand this assuming that $c_1\ll\mu\ll c_2$.
In the channel corresponding to the small eigenvalue, this gives the diagonal interaction 
\beq
\begin{split}
\hat{V}_\mathrm{u}&\equiv [f_\mathrm{u}^{-1}-f^{\ }_\mathrm{v}\,f_\mathrm{m}^{-1}f_\mathrm{m}^{*-1}]^{-1}\\
&=\left[-1-\sum_{n,m}c_{nm}^{(\mathrm{u})}\,\mu^{2n+2m-1}\,\hat p^{2n}\,
\hat \delta^{2m}
+\frac{\mu^3}{c_2}\left|c^{(\mathrm{m})}_{10}\hat p^2+c^{(\mathrm{m})}_{01}\hat\delta^2\right|^2
+\cdots\right]^{-1}\\
&=-1+\sum_{n,m}c_{nm}^{(\mathrm{u})}\,\mu^{2n+2m-1}\,\hat p^{2n}\,
\hat \delta^{2m}+\cdots,
\end{split}
\eeq
where terms suppressed by additional powers of $\mu/c_2$ or $c_1/\mu$ have been omitted.
The dominant terms multiplying each product of powers of $p$ and $\delta$ have the same
power counting as in  the effective-range expansion, that is, enhanced by two orders
over simple dimensional analysis. 

In contrast, the diagonal interaction in the other channel has the expansion
\beq
\begin{split}
\hat{V}_\mathrm{v}&\equiv [f_\mathrm{v}^{-1}-f^{\ }_\mathrm{u}\,f_\mathrm{m}^{-1}f_\mathrm{m}^{*-1}]^{-1}\\
&=\left[-\sum_{n,m}c_{nm}^{(\mathrm{v})}\,\mu^{2n+2m-1}\,\hat p^{2n}\,\hat \delta^{2m}-1
+\mu^2\left|c^{(\mathrm{m})}_{10}\hat p^2+c^{(\mathrm{m})}_{01}\hat\delta^2\right|^2+\cdots\right]^{-1}\\
&=\mu\left[-\sum_{n,m}c_{nm}^{(\mathrm{v})}\,\mu^{2n+2m}\,\hat p^{2n}\,\hat \delta^{2m}
+\mu\left(-1+\mu^2\left|c^{(\mathrm{m})}_{10}\hat p^2+c^{(\mathrm{m})}_{01}\hat \delta^2\right|^2
+\cdots\right)\right]^{-1}\\
&=-\frac{\mu}{c_2}+\frac{1}{c_2^2}\,{\sum_{n,m}}^{\prime}c_{nm}^{(\mathrm{v})}\,
\mu^{2n+2m+1}\,\hat p^{2n}\,\hat \delta^{2m}+\cdots,
\end{split}
\eeq
where the prime indicates the omission of the term with $n=m=0$. This is just the 
standard Weinberg power counting, as expected for a channel with no low-energy bound or
virtual state. The renormalisation in this channel is done perturbatively; the terms (suppressed in the equation above)
necessary for this purpose do not affect the power counting. Finally, the off-diagonal interactions can be expanded as
\beq
\begin{split}
\hat{V}_\mathrm{m}&\equiv -f_\mathrm{m}^{-1}f^{\ }_\mathrm{v}[f_\mathrm{u}^{-1}-f^{\ }_\mathrm{v}\,f_\mathrm{m}^{-1}f_\mathrm{m}^{*-1}]^{-1}\\
&=\frac{1}{c_2}\sum_{n,m}c_{nm}^{(\mathrm{m})}\,\mu^{2n+2m}\,\hat p^{2n}\,\hat \delta^{2m}+\cdots.
\end{split}
\eeq
The terms in these are enhanced by one power compared to simple dimensional analysis.

In the corresponding unscaled potential, the RG eigenvalues translate to powers of 
low-energy scales, generically denoted by $Q$, a term that scales as $\mu^\nu$ being
of order $Q^d$ with $d=\nu-1$ \cite{Birse:1998dk}. The small scales here are $p$ and 
$\delta$, and also the subtraction scale $\mu$, which can be regarded as the highest 
acceptable momentum scale in our EFT. However, terms with different powers of 
$\delta$ cannot be distinguished in practice and so they can be grouped 
together to form observables, such as the scattering length in the 
strongly-interacting channel,
\beq
\frac{1}{a_1}=-c_1+{\cal O}(\delta^2).
\eeq 

The unscaled inverse potential corresponding to Eq.~\eqref{eqn:Vinvgen} can then be 
expressed in the form
\beq
\mathbf{V}^{-1}=-\frac{1}{2\pi}\mathbf{M}^{1/2}\left[\left(\mu-\frac{1}{a_1}
+\frac{r_0}{2}p'^2\right)\mathbf{P}_\mathrm{u}+\left(\mu-\frac{1}{a_2}\right)\mathbf{P}_\mathrm{v}
+\frac{r_1}{2}\,p'^2\left(\mathbf{P}_\mathrm{m}+\mathbf{P}_\mathrm{m}^\dagger\right)+\cdots
\right]\mathbf{M}^{1/2},
\label{eqn:VNNLOINV}
\eeq
where we have expanded in powers of energy around the higher threshold, with
$p'\equiv p_2$, and kept terms to NNLO (order $Q^3$) in the power counting defined above.
In this counting the large scattering length $a_1$ is of order $Q^{-1}$, while 
all other coefficients are of natural size, that is, of order $Q^0$. We have assumed 
that no other channels are open and hence we have taken the mixing parameter $r_1$ 
to be real.

The Lippmann-Schwinger equation is conveniently written in the form
\beq
\mathbf{T}^{-1}=\mathbf{V}^{-1}-\mathbf{J},
\label{eqn:LSE_inv}
\eeq
where the terms proportional to $\mu$ in the loop integrals can immediately be seen 
to cancel against the corresponding terms in $\mathbf{V}^{-1}$. The resulting
expression for $\mathbf{T}^{-1}$ can then be inverted and expanded in the same manner 
as used above to construct the potential $\mathbf{V}$. The resulting expression for the scattering matrix is, to NNLO, 
\beq
\begin{split}
\mathbf{T}^\mathrm{NNLO}=-2\pi\,\mathbf{M}^{-1/2}
&\Biggl\{\biggl[
-\,\frac{1}{a_1}+\frac{r_0}{2}\,p'^2-\mathrm{i}p_\mathrm{u}-a_2\left(\left[1-\mathrm{i}a_2p_\mathrm{v}\right]p_\mathrm{m}^2
-\mathrm{i}r_1\,p'^2p_\mathrm{m}\right)\biggr]^{-1}\mathbf{P}_\mathrm{u}\\
&\ -a_2\biggl[1-\mathrm{i}a_2p_\mathrm{v}
+a_2p_\mathrm{m}^2\left(-\frac{1}{a_1}-\mathrm{i}p_\mathrm{u}\right)^{-1}\biggr]\mathbf{P}_\mathrm{v}\\
&\ +a_2\left[\mathrm{i}p_\mathrm{m} \left(\left[-\frac{1}{a_1}-\mathrm{i}p_\mathrm{u}\right]\left[1+\mathrm{i}a_2p_\mathrm{v}\right]+\frac{r_0}{2}p'^2-a_2p_\mathrm{m}^2\right)^{-1}\right.\\
&\ \left.\hphantom{+a_2\biggl[}+\frac{r_1}{2}\,p'^2\left(-\frac{1}{a_1}-\mathrm{i}p_\mathrm{u}\right)^{-1}
\right]
(\mathbf{P}_\mathrm{m}+\mathbf{P}_\mathrm{m}^\dagger)\Biggr\}\mathbf{M}^{-1/2},
\end{split}
\label{eqn:TNNLO_sr}
\eeq
where we defined the momentum variables
\beq
p_\mathrm{u}=p\cos^2\phi+p'\sin^2\phi,\ p_\mathrm{v}=p\sin^2\phi+p'\cos^2\phi,\ p_\mathrm{m}=(p-p')\sin\phi\cos\phi.
\eeq
Here we have chosen to leave the diagonal amplitude in the strongly-interacting channel,
as well as parts of the amplitude in the mixing channel, in the form of an effective-range
expansion but, if desired, these terms could also be expanded to order $Q^3$.

\subsection{Short-range plus Coulomb}

With some modifications, the analysis above can also be applied in the presence of the Coulomb interaction in either or both of 
the channels. This introduces two new low-energy scales, the inverse Bohr radii for the 
two channels, $\kappa_{1,2}$. The Lippmann-Schwinger equation takes the form
\beq
\mathbf{T}_C(p,\delta,\kappa_{1,2})=\mathbf{V}(p,\delta,\kappa_{1,2},\mu)
+\mathbf{V}(p,\delta,\kappa_{1,2},\mu)\mathbf{J}(p,\delta,\kappa_{1,2},\mu)
\mathbf{T}_C(p,\delta,\kappa_{1,2}).
\eeq
The individual loop integrals in $\mathbf{J}(p,\delta,\kappa_{1,2},\mu)$ 
are given by
\beq
J_i^\mathrm{MS}(p,\delta,\kappa,\mu)=-\,\frac{M_i}{2\pi}\,
\left(\mu-2\kappa_i\left\{\ln\frac{2\mu\sqrt{\pi}}{\kappa_i}+1
-\frac{3}{2}\gamma_E\right\}+2\kappa_i\left[h(\eta_i)+\mathrm{i}\,\frac{C^2_{\eta_i}}{2\eta_i}\right]\right),
\eeq
where $\eta_i=\kappa_i/p_i$, $h(z)=\mathrm{Re}\, \psi(\mathrm{i}z)-\ln z$ ($\psi$ 
denoting the logarithmic derivative of Euler's gamma function), $C_{\eta_i}^2=2\pi\eta_i/(\exp 2\pi\eta_i-1)$ are the Sommerfeld factors,
and $\gamma_E=-0.5772\dots$ is the Euler-Mascheroni constant. To get this, we have used PDS 
for the linear divergence and minimal subtraction for the logarithmic one induced by 
the Coulomb potential. 

These logarithmic divergences in the basic loop integrals generate logarithmic dependences 
of the potential on $\mu$. The resulting general solution to the RG equation can be written
\beq
\hat{\mathbf{V}}(\hat p,\hat\delta,\hat\kappa,\mu)^{-1}=-\mathbf{I}
+\hat{\boldsymbol{\kappa}}\, 2\ln\frac{\mu}{\lambda}-\sum_{n,m,l}\mathbf{C}_{nml}
\,\mu^{2n+2m+l-1}\,\hat p^{2n}\,\hat \delta^{2m}\hat{\kappa}^l,
\label{eqn:VinvgenC}
\eeq
where $\lambda$ is an arbitrary energy scale, and $\hat{\boldsymbol{\kappa}}$ is the 
diagonal matrix of the (rescaled) inverse Bohr radii. Note that $\mathbf{C}_{001}$ 
must depend on $\lambda$ so as to cancel the explicit dependence on $\lambda$ in
the second term. 

As in the single-channel case, the only true fixed-point solution is the trivial one
$\hat{\mathbf{V}}=0$ \cite{Barford:2002je,ab08,ab08_1}. This is because both of the nontrivial 
fixed points found above become unstable as a result of the logarithmic evolution induced by 
the Coulomb potential. Nonetheless, we can still construct solutions that describe systems 
with either one or two bound states that are close to threshold on the energy scales 
of interest. Again, it is the expansion of the potential around the solution with a 
single such state that is of more interest since it requires only one ``fine tuning",
corresponding to one unnaturally small eigenvalue of $\mathbf{C}_{000}$. The power 
counting for this expansion is similar to that derived above. A minor modification
is the presence of the short-distance interactions proportional to powers of $\kappa$, 
although these cannot be disentangled phenomenologically from purely strong-interaction 
terms \cite{kr99,Ando:2007fh,Barford:2002je}. 

The corresponding scattering matrix has a very similar form to that in the 
purely short-range case. Expanded to NNLO, it can be written
\beq
\begin{split}
\mathbf{T}^\mathrm{NNLO}=-2\pi\,\mathbf{M}^{-1/2}
&\Biggl\{\biggl[
-\,\frac{1}{a_1}+\frac{r_0}{2}\,p'^2-j_\mathrm{u}(p)+a_2\left(j_\mathrm{m}^2(p)\left[1-a_2 j_\mathrm{v}(p)\right]+r_1\,p'^2\, j_\mathrm{m}(p)\right)
\biggr]^{-1}\mathbf{P}_1\\
&\ -a_2\left[1-a_2 j_\mathrm{v}(p)-a_2 j_\mathrm{m}^2(p)\left(-\frac{1}{a_1}-j_\mathrm{u}(p)\right)^{-1}\right]\mathbf{P}_2\\
&\ +a_2\left[j_\mathrm{m}(p)\left(\left[-\frac{1}{a_1}-j_\mathrm{u}(p)\right]\left[1+a_2 j_\mathrm{v}(p)\right]+\frac{r_0}{2}p'^2+a_2 j_\mathrm{m}^2(p)\right)^{-1}\right.\\
&\ \hphantom{+a_2\Biggl[}
\left.+\frac{r_1}{2}\,p'^2\left(-\frac{1}{a_1}-j_\mathrm{u}(p)\right)^{-1}\right]
(\mathbf{P}_3+\mathbf{P}_4)\Biggr\}\mathbf{M}^{-1/2},
\end{split}
\label{eqn:TNNLO_co}
\eeq
where we have used the fact that the linear and logarithmic dependences on $\mu$ cancel
in the Lippmann-Schwinger equation Eq.~\eqref{eqn:LSE_inv}. This allows us to express 
the result in terms of the finite parts of the loop integrals in the presence of 
the Coulomb potential,
\beq
j_i(p)=\kappa_i\left(2h(\eta_i)+\mathrm{i}\,\frac{C^2_{\eta_i}}{\eta_i}\right),
\eeq
where $i=1,2$ denote the physical channel, and the relevant combinations of $j_{i}(p)$ are
defined analogously to the purely strong-interaction case: 
\beq
\begin{split}
j_\mathrm{u}(p)=& j_1(p)\cos^2\phi+j_2(p)\sin^2\phi,\\ j_\mathrm{v}(p)=& j_1(p)\sin^2\phi+j_2(p)\cos^2\phi,\\ j_\mathrm{m}(p)=& \left(j_1(p)-j_2(p)\right)\sin\phi\cos\phi.
\end{split}
\eeq
For the NLO expression, the terms with $r_1$ should be omitted, as well as the terms
proportional to $a_2^2$ in the expansion, and the term with $r_0$ in the mixing channel;
for the LO one, the terms with $a_2$ and the remaining term with $r_0$ should also be left out.

\section{$\mbox{\textit{p}}+^7\mbox{Li}\leftrightarrow \mbox{\textit{n}}+{^7}\mbox{Be}$
 Coupled Channels}\label{sec:CC}
\subsection{EFT description of the data}

To illustrate the application of an EFT based on the power counting outlined above,
we examine here the $p+^7\mbox{Li} \leftrightarrow n+{^7}\mbox{Be}$ coupled-channel 
system. The $n+{^7}\mbox{Be}$ threshold lies at $\Delta=1.6442$~MeV and there is a 
$J^P=2^-$ excited state of ${^8}$Be within a few keV of this threshold 
\cite{JPhysG.29.395,PhysRevC.37.917}. This state manifests itself as a prominent peak in 
the $p+^7\mbox{Li}$ elastic scattering cross sections~\cite{NuclPhysA.206.353} and 
a very large cross section, $\sigma\simeq 38.4\times 10^3$~b \cite{PhysRevC.37.917}, for
the reaction ${^7}\mathrm{Be}(n,p){^7}\mathrm{Li}$ at low (thermal) energies.
Together with the absence of any unusual features in the ${^5}S_2$ phase shift below 
the $n+{^7}\mbox{Be}$ threshold \cite{NuclPhysA.206.353}, this makes the system an 
ideal candidate to be studied with our approach.

The system has two important low-energy scales: the momentum scale corresponding to the
 difference between the thresholds, $\delta=51.95$ MeV, and the inverse Bohr radius 
for the $p+^7\mbox{Li}$ channel, $\kappa_1\equiv\kappa=17.96$ MeV. These, together with
the on-shell relative momentum, provide the expansion parameters of the EFT.

The scattering matrix  for this system is given by Eq.~\eqref{eqn:TNNLO_co}, 
with the simplification that $j_2(p')=\mathrm{i}p'$ since the neutron in the 
second channel is uncharged.

We now confront our EFT with the available data in the $p+{^7}\mathrm{Li}$ and $n+{^7}\mathrm{Be}$ channels. This data consists of
the ${^5\!}S_2$ phase shift, $\delta_0$, in ${^7}\mathrm{Li}+p$ elastic scattering,
where results of partial-wave analyses and $R$-matrix fits exist~\cite{NuclPhysA.206.353}, and the cross section, $\sigma_{np}$,
of the charge exchange $\mathrm{{^7}Be}(n,p)\mathrm{{^7}Li}$ reaction, from high quality measurements
of Ref.~\cite{PhysRevC.37.917}, and the $R$-matrix fit of Ref.~\cite{JPhysG.29.395}. These quantities are related to the
corresponding elements of the $\mathbf{T}$ matrix via
\bea
\frac{\rho e^{2\mathrm{i}\delta_0}-1}{2\mathrm{i} p}&=&-\frac{M_1}{2\pi}\, T_{11}C^2_\eta,\\
\sigma_{np}&=&\frac{4\pi(2J+1)}{(2s_1+1)(2s_2+1)}\frac{p}{p'}\frac{M_1 M_2}{4\pi^2}\left|-C_\eta\,T_{21}\right|^2,
\eea
where $J=2$ is the total momentum in the partial wave considered, $s_1=3/2$, $s_2=1/2$ are the spins of initial particles, $\eta=\kappa/p$, and the Sommerfeld factors
arise due to the wave functions of the corresponding Coulomb scattering states. Here, $\rho$ is the measure of elasticity in $p+{^7}$Li channel,
equal to one below $n+{^7}\mathrm{Be}$ threshold.

Before we proceed to the description of our results, a few words on our fitting procedure are in order.
 We fit NLO and NNLO results to the phase shift and the reduced $(n,p)$ cross section, $\sigma_{np}^\mathrm{red}=\sigma_{np}\sqrt{E-\Delta}$,
while the LO parameters are extracted from the threshold values of the phase shift and the reduced cross section, rather
than fitted over the whole region (such LO fits give similar values of parameters, and reproduce the observables with comparable quality).
The results of phase analysis of Ref.~\cite{NuclPhysA.206.353} do not provide an uncertainty
for the phase shift. We therefore assign an estimated uniform uncertainty of $\pm 5^\circ$ to the data points. This estimate
is based on the scatter of the data points and is used in the chi-square fitting function.
We do not aim to reproduce the features of $\delta_0$ at energies above the neutron threshold, for the reasons that will be discussed below.
As to the cross section, we take the experimental data and uncertainties on the total cross section from Ref.~\cite{PhysRevC.37.917},
appropriately rescaled to obtain the reduced cross section. Except at LO, we performed fits of two kinds  --- in one case,
we fitted the expanded $\mathbf{T}$ matrix, as in Eq.~\eqref{eqn:TNNLO_co}, whereas the other fit employed the exact $\mathbf{T}$ matrix,
obtained by inverting the Lippmann-Schwinger equation with $\mathbf{V}^{-1}$ at a given order. This form preserves unitarity exactly.
In the fits of the expanded $\mathbf{T}$ matrix, we also investigated the role of constraining them to satisfy unitarity below the threshold,
which was done by imposing a constraint that $\rho$ be close to one.
At each order we fit all the parameters that enter $\mathbf{V}^{-1}$ at that order.

Before presenting our results, we should comment on some of the numerical details of the different fits. At NLO, all the versions of our fits resulted in very similar
values for the low-energy parameters, except for $a_2$, which was very small in all cases, although its values could differ by a factor of two. 
At NNLO, we encountered two kinds of numerical difficulties. The first was that some fits, namely those at NNLO with the expanded $\mathbf{T}$ matrix but without the unitarity constraint,
resulted in far too high values of the reduced cross section at higher energies, $E-\Delta>0.1$~MeV, despite still describing the available data well.
In this region there are no direct experimental data on the cross section for this partial wave, as a result of other partial waves, especially $J^P=3^+$,
becoming important. We therefore used the results of the $R$-matrix fit of Ref.~\cite{JPhysG.29.395} as reference points at these energies, in order to filter out fits
that led to unnaturally large value of the cross section at higher energies. The second difficulty was related to the fact that already at NLO we reproduce
the available data quite well, and hence adding a new parameter, $r_1$, results in a very broad minimum
of the fitting function at NNLO. This indicates that the available data is not enough in order to constrain the scattering parameters beyond NLO, and, as a result,
the variations of the scattering parameters between the different versions of fits grow significantly compared to NLO.

\begin{figure}
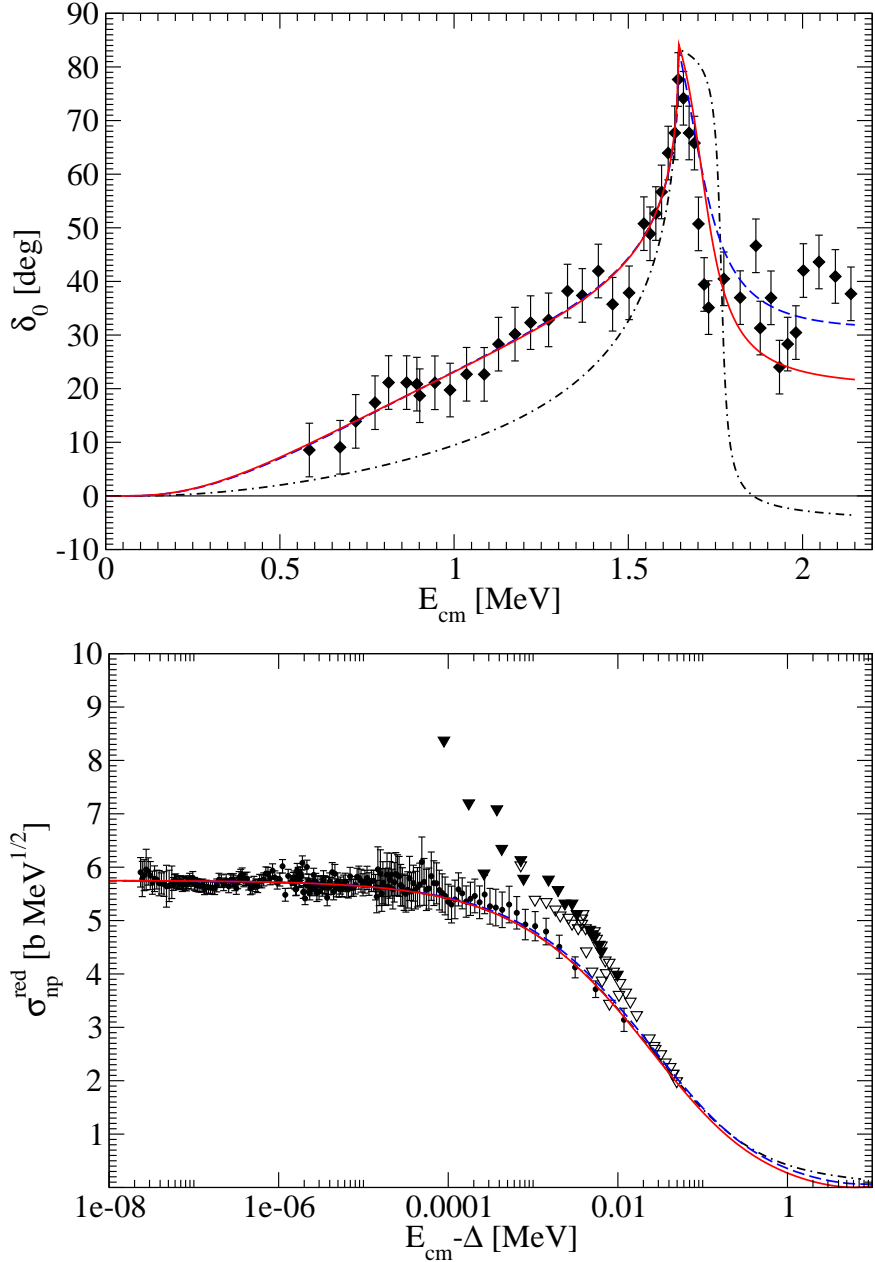

\begin{tabular}{c}
\epsfig{file=phase_orders_exp.eps,width=0.7\textwidth,clip=}\\
\epsfig{file=xs_orders_exp.eps,width=0.7\textwidth,clip=}
\end{tabular}
\caption{Upper panel: phase shift at different orders, fits of expanded $\mathbf{T}$ matrix without unitarity constraint.
Dash-dotted curve --- LO, dashed curve --- NLO, solid curve --- NNLO. Data: diamonds --- phase analysis
from Ref.~\cite{NuclPhysA.206.353} (digitized; the data uncertainties are as described in the text). The dashed and the solid curves are on top of each other below the neutron threshold.
Note that only the phase shift data below the threshold are included in the fits, as explained in the text.\\
Lower panel: reduced cross section of the reaction ${^7}$Be$(n,p){^7}$Li. The curves are as in the upper panel.
Data points: 
circles --- data from Ref.~\cite{PhysRevC.37.917},
triangles --- derived from the near-threshold data (digitized) of Ref.~\cite{PhysRev.109.105} on the crossed
reaction, ${^7}$Li$(p,n){^7}$Be, hollow and filled
triangles correspond to targets a and b of that reference, in order.
}\label{fig:Discussion}
\end{figure}

In Table~\ref{tab:param} we show the values of the scattering parameters that resulted from our fits, using the expanded $\mathbf{T}$ matrix without the unitarity constraint.
At LO, taking the value of ${^5\!}S_2$ phase shift at the neutron threshold $83.3^\circ$ and the value of the reduced $(n,p)$ cross section $5.75$ b MeV${^{1/2}}$ gives
\bea
a_1&=&-17.76~\mathrm{fm},\\
\phi&=&46.63^\circ.
\eea
One can see from Table~\ref{tab:param} that the low-energy parameters $a_2$, $r_0$ and $r_1$ are of natural size
for an effective theory with an underlying length scale $\sim 2$~fm.
The values of $a_1$ and $\phi$ change slightly when one goes to NLO and NNLO. 
Our results for the phase shift and the reduced $(n,p)$ cross section are shown in Fig.~\ref{fig:Discussion}. 
Surprisingly, the LO description of $\sigma_{np}^\mathrm{red}$ is already very good up to energies $\sim 0.5$~MeV
above neutron threshold. In fact, all three curves, LO, NLO, and NNLO, differ only slightly and only at higher energies
(even though the $p+{^7}$Li phase shift $\delta_0$ at these energies is not well described at LO). Furthermore, we observed that fitting only
the threshold value of $\sigma_{np}^\mathrm{red}$ without paying attention to the details of its energy dependence, produces, as a rule, fits of good quality.
In contrast to this, the behaviour of the phase shift above the neutron threshold is very unstable at LO --- slight changes of threshold values
of $\delta_0$ and $\sigma^\mathrm{red}_{np}$ can lead to very different behaviour of $\delta_0$. This is due to
the closeness of $\delta_0$ to $90^\circ$ at threshold and a strong inelasticity in the proton channel just
above the neutron threshold. This instability, however, was not observed in NLO and NNLO fits.

The phase shift below the neutron threshold is reproduced quite nicely already at NLO, and its convergence can be seen from Fig.~\ref{fig:Discussion}.
At the same time, the slope of $\delta_0$ above threshold is not reproduced very well.
We believe this to be due to the inconsistency between the data on the reaction ${^7}\mathrm{Be}(n,p){^7}$Li of Ref.~\cite{PhysRevC.37.917}
used in our fits and the (older) data on the inverse reaction, ${^7}\mathrm{Li}(p,n){^7}$Be, measured in Ref.~\cite{PhysRev.109.105}
and used in the phase-shift analysis of Ref.~\cite{NuclPhysA.206.353}. The data of Ref.~\cite{PhysRev.109.105} are significantly higher
than those of Ref.~\cite{PhysRevC.37.917}, as also illustrated in Fig.~\ref{fig:Discussion}. This means that the inelasticity
in the $p+{^7}$Li channel above the neutron threshold was larger in that analysis than in our fits, which could explain the
difference between the slopes. This inconsistency between the phase shifts above threshold and the data on $\sigma_{np}$ led us
to discard the phase shifts above threshold from our fits, as mentioned above.
On the other hand, since the inelasticities in $p+{^7}$Li scattering are small below the neutron threshold, we have no reason
to question the reliability of the results of the phase-shift analysis at these energies.

\begin{table}
\begin{tabular}{||c|c|c|c|c|c||}
\hline
Order      & $a_1$ [fm]         & $\phi$                       & $a_2$ [fm]        & $r_0$ [fm]        & $r_1$ [fm]\\
\hline
LO         & $-17.76$           & $\ 46.63^\circ$    & $-$               & $-$               & $-$               \\
\hline
NLO        & $-19.37$           & $\ 51.82^\circ$    & $-1.96$           & $3.79$            & $-$               \\
\hline
NNLO       & $-19.11$           & $\ 50.45^\circ$    & $-1.07$           & $2.58$            & $-5.42$           \\
\hline
\end{tabular}
\caption{Scattering parameters at different orders, resulting from the fits of the expanded $\mathbf{T}$ matrix, Eq.~(\ref{eqn:TNNLO_co}), 
without the unitarity constraint, as described in the text.}
\label{tab:param}
\end{table}

\subsection{The structure of the $2^-$ ${^8}$Be state near the neutron threshold}

\begin{figure}
\epsfig{file=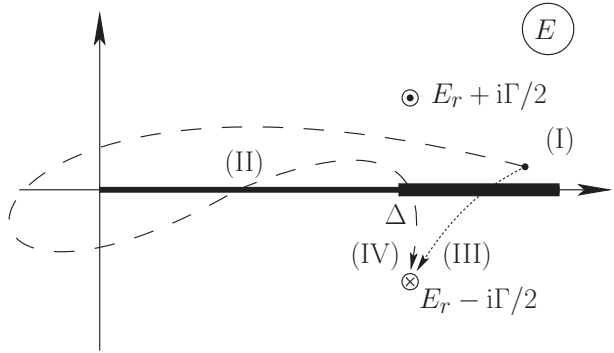,width=0.5\textwidth,clip=}
\caption{Poles and zeros of the $\mathbf{S}$ matrix in the energy plane. Dotted line --- a path from a point
near the physical region, showed by a dot, to sheet III; dashed line --- a path from the physical
region to sheet IV. Roman numbers in brackets along the paths indicate the corresponding change of the sheets.
The location of the pole in sheet IV, coinciding with the location of a zero
of $S_{22}$ on the physical sheet, is denoted by the circled cross. The location of the conjugated zero of
$S_{22}$ on the physical sheet is shown by the circled dot.
}\label{fig:Pole}
\end{figure}

Here we would like to discuss the structure of the $2^-$ ${^8}$Be state near the neutron threshold.
In our calculation, it appears as a pole in the $\mathbf{S}$ matrix located right above the threshold,
at $E=E_r-\mathrm{i}\Gamma/2=1.71-\mathrm{i}0.06$ MeV, giving the total width $\Gamma=0.12$ MeV.
These numbers correspond to the NNLO parameters from Table~\ref{tab:param}. Note that due to the
expansion, the different channels ($\mathbf{u}$, $\mathbf{v}$, and the mixing channel) of the $\mathbf{T}$ matrix, Eq.~\eqref{eqn:TNNLO_co},
have poles at slightly different positions, which would not be the case for the exact $\mathbf{T}$ matrix.
The numbers quoted for the pole position correspond to the pole in the $\mathbf{u}$ channel. The errors
introduced by the expansion are, however, not relevant for the following discussion.
In the commonly used notation~\cite{PhysRev.134.B1307},
where the four sheets of the energy Riemann surface are defined as
\begin{itemize}
\item $\mathrm{Im}\, p_1>0,\ \mathrm{Im}\, p_2>0$: sheet I (physical);
\item $\mathrm{Im}\, p_1<0,\ \mathrm{Im}\, p_2>0$: sheet II;
\item $\mathrm{Im}\, p_1<0,\ \mathrm{Im}\, p_2<0$: sheet III;
\item $\mathrm{Im}\, p_1>0,\ \mathrm{Im}\, p_2<0$: sheet IV,
\end{itemize}
this pole is located on sheet IV. A pole of this kind, being an analogue of a single channel virtual state, was termed
a ``shadow resonance'' by Eden and Taylor~\cite{Eden:1964zz}. It should be distinguished from a ``regular'' Breit-Wigner
resonance located on sheet III. The most important difference
between a shadow resonance and a Breit-Wigner resonance is that, while the latter is accessible
immediately from the physical region, i.e.\ the upper edge of the cut on the physical sheet,
by moving down across the cut at $\mathrm{Re}\,E>\Delta$, the former can be accessed from the physical region only 
by moving around the zero-energy threshold, and is thus rather far away from the physical region.
These different paths, corresponding to poles on sheets III and IV, are illustrated in Fig.~\ref{fig:Pole}.
A very well-known consequence of a Breit-Wigner pole, lying on sheet III close to the physical region,
is the expansion of the $\mathbf{S}$ matrix close to the pole, having the form 
\bea
\mathbf{S}=\mathbf{I}
%\mathbf{1}
-\frac{\mathrm{i}\mathbf{A}}{E-E_r+\mathrm{i}\Gamma/2},
\label{eqn:SmatRes}
\eea
where $\mathbf{A}$, the residue of the $\mathbf{S}$ matrix, is a Hermitian rank one matrix, satisfying
\beq
\sum_i\left|A_{ii}\right|\equiv\sum_i\Gamma_i=\Gamma,
\label{eqn:Amat}
\eeq
where the quantities $\Gamma_i$ are identified with the partial widths corresponding to the $i^\mathrm{th}$ channel.
These properties of $\mathbf{A}$ are dictated by the unitarity of the $\mathbf{S}$ matrix in the physical region.
A pole on sheet IV, on the contrary, does not allow for a similar expansion of the $\mathbf{S}$ matrix 
in the physical region, due to the large distance to the pole. As a result, the residue, $\mathbf{A}$, is no
longer constrained to be Hermitian. The condition of Eq.~(\ref{eqn:Amat}) is broken in this case as well:
although the quantities $\left|A_{ii}\right|=\Gamma_i$ can still be identified with the partial widths
(see, e.g., Refs.~\cite{PhysRevC.37.917,JPhysG.29.395,Hale:1987zz}),
they do not add up to the total width. In our case $\Gamma_p=2.47$ MeV, $\Gamma_n=0.34$ MeV,
corresponding to the so-called strength of the resonance $\sum_i \Gamma_i/\Gamma=22.89$.

It is interesting to note that the pole, despite being located far away from the physical region,
still crucially affects the $\mathbf{S}$ matrix and the observables close to the neutron threshold,
via the zeros of certain elements of the $\mathbf{S}$ matrix and unitarity
(see Ref.~\cite{Hale:1987zz} for a discussion of an analogous situation in the coupled system
$d(t,n){^4}$He). Namely, associated with the pole on sheet IV, there is a zero of one of
the elements of the $\mathbf{S}$ matrix, $S_{22}$, located at the same energy on sheet I,
as discussed in Ref.~\cite{Eden:1964zz}. Although this zero is on sheet I,
it is still located rather far away from the physical region. However, it can be easily shown
that there is another zero of $S_{22}$, located at the complex-conjugate point, $E=E_r+\mathrm{i}\Gamma/2$ on sheet I.
This zero is, in contrast to the pole and the zero discussed above, located very close
to the physical region, as is also illustrated in Fig.~\ref{fig:Pole}. The nearby zero forces $S_{22}$
(and, through unitarity, also $S_{11}$ above the neutron threshold) be small at $E\simeq E_r$,
while the off-diagonal element, $S_{12}$, approaches the unitary limit. These features are seen in the observables.

\section{Discussion}\label{sec:Conclusion}

In this paper, we have used an RG analysis to elucidate the power counting needed 
to organise an EFT describing a system of coupled channels with a single 
low-energy state close to one the thresholds. Such a system has, as its low-energy
scales, the momenta corresponding to the various thresholds, as well as the 
on-shell momentum. These provide the expansion parameters of the corresponding EFT.

The RG analysis of the two-channel case leads to the identification of three 
fixed points. One of these is just the trivial one. The expansion around it can
be used to analyse systems that are weakly interacting in all channels. A second
has two low-energy bound or virtual states and so both channels are strongly
interacting at low energies. Lastly, of most interest in practice is one with
a single bound or virtual state. Here, one linear combination of the two 
asymptotic channels is strongly interacting, the other weakly so.

The power counting for the strongly interacting channel near that third fixed point 
is that of an effective-range expansion. As in the corresponding single-channel case,
terms in the effective potential are promoted by two orders relative to simple
dimensional analysis. The leading, energy-independent term generates large 
scattering length. It is a relevant perturbation and so must be treated to all 
orders. All other terms in the potential can be treated as perturbations.

All terms in the weakly interacting channel are of natural strength and so can
be organised according to simple dimensional analysis. The off-diagonal interactions
that couple the two channels are promoted by one order. The leading one of
these is then a marginal perturbation, but it can be absorbed into the mixing
angle that defines the strongly interacting channel.

The resulting expansion can be applied to ``halo" EFTs that describe 
nuclear systems with weakly bound low-energy states~\cite{bhvk02,bhvk03,hhvk08,rh11,hp11}.
It may also be applicable to some of the states seen close to thresholds in the quarkonium systems~\cite{hkn11}.

In general the channels in these examples involve charged particles and so
we have extended our RG analysis to include Coulomb potentials between the 
particles. As in similar analyses of the single-channel case, these potentials
are not sufficiently singular to substantially change the power counting. One 
obvious effect is to provide additional low-energy scales: the inverse Bohr 
radii for the various channels.

In the strongly interacting channel, Coulomb forces introduce a marginal interaction.
As a result, there is no longer a true fixed point but instead 
a logarithmically evolving RG trajectory. Theories near this trajectory can still be organised 
according to a modified effective-range expansion in their strongly interacting 
channel. The weakly interacting channel is still natural, while off-diagonal 
interactions are again promoted by one order.

To demonstrate the use of the resulting EFT, we have applied it to an ideal 
test case: the coupled $p+{^7}$Li and $n+{^7}$Be channels. They couple to a 
$J^P=2^-$ state of $^8$Be which lies within a few keV of the $n+{^7}\mbox{Be}$
threshold.

At NLO we get a good description of the available  $p+{^7}$Li phase shift and 
the $\mathrm{{^7}Be}(n,p)\mathrm{{^7}Li}$ reaction cross section using four parameters.
At NNLO there is one further parameter, and although we are able to get fits of similar
quality, we find evidence that the available data are not sufficient to determine
all of the parameters at this order.
The sizes of the perturbative terms are consistent with an underlying scale of 
the order of $50-100$~MeV. The scattering length in the strongly interacting
channel is about 0.1~MeV$^{-1}$, making it unnaturally large on this scale. The
differences between the paramaters from fits at LO, NLO and NNLO are also as
expected.

As in previous analyses of this system, we find that the $J^P=2^-$ state of 
$^8$Be is described by a pole on the fourth sheet. This lies far from the physical 
sheet so it is not a resonance but rather the multi-channel analogue of a 
virtual state. It makes its presence known through the cusp in the $p+{^7}$Li
phase-shift at the $n+{^7}$Be threshold and the very large cross section for 
$\mathrm{{^7}Be}(n,p)\mathrm{{^7}Li}$ at low energies.

The example demonstrates the viability of this EFT approach for analysing 
coupled-channel systems with low-energy bound or virtual states. It may provide 
a more systematic alternative to the $R$-matrix method which is widely used 
in studies of such systems. It can also be applied to similar states in the 
quarkonium systems. 

\section*{Acknowledgments}

We thank C.~Hanhart and A.~E.~Kudryavtsev for their useful comments.
MCB is grateful to G.~Hale for the discussions that prompted this investigation, 
and to the Institute for Nuclear Theory for providing the opportunity for these. 
This work was supported by the UK STFC under grant ST/F012047/1.

\end{document}